\documentclass{article}
\usepackage[utf8]{inputenc}
\usepackage{graphicx}
\usepackage{multirow}
\usepackage{amsmath}				
\usepackage[colorlinks=true,linkcolor=red,citecolor=blue]{hyperref}
\usepackage[font=small,labelfont=bf]{caption}
\usepackage{cite}
\usepackage{textcomp}
\usepackage{arxiv}
\usepackage[T1]{fontenc}    
\usepackage{url}            
\usepackage{booktabs}       
\usepackage{amsfonts}       
\usepackage{nicefrac}       
\usepackage{microtype}      
\usepackage{lipsum}					
\usepackage{amssymb}
\usepackage[shortlabels]{enumitem}
\usepackage{verbatim}
\usepackage{float}
\usepackage{newfloat}

\title{COVID-19 Data Analysis and Forecasting: Algeria and the World}

\date{May 19, 2020}

\author{Sami Belkacem \\
				Department of Computer Science\\
				USTHB University\\
				Bab-Ezzouar, Algiers, Algeria\\
				\texttt{s.belkacem@usthb.dz} \\ }

\let\oldmaketitle\maketitle
\renewcommand{\maketitle}{\oldmaketitle\setcounter{footnote}{0}}

\begin{document}
\maketitle

\begin{abstract}
The novel coronavirus disease 2019 COVID-19 has been leading the world into a prominent crisis. As of May 19, 2020, the virus had spread to 215 countries with more than 4,622,001 confirmed cases and 311,916 reported deaths worldwide, including Algeria with 7201 cases and 555 deaths. Analyze and forecast COVID-19 cases and deaths growth could be useful in many ways, governments could estimate medical equipment and take appropriate policy responses, and experts could approximate the peak and the end of the disease. In this work, we first train a time series \textit{Prophet} model to analyze and forecast the number of COVID-19 cases and deaths in Algeria based on the previously reported numbers. Then, to better understand the spread and the properties of the COVID-19, we include external factors that may contribute to accelerate/slow the spread of the virus, construct a dataset from reliable sources, and conduct a large-scale data analysis considering 82 countries worldwide. The evaluation results show that the time series \textit{Prophet} model accurately predicts the number of cases and deaths in Algeria with low RMSE scores of 218.87 and 4.79 respectively, while the forecast suggests that the total number of cases and deaths are expected to increase in the coming weeks. Moreover, the worldwide data-driven analysis reveals several correlations between the increase/decrease in the number of cases and deaths and external factors that may contribute to accelerate/slow the spread of the virus such as geographic, climatic, health, economic, and demographic factors.
\end{abstract}

\section{Introduction}\label{sec1}
The novel coronavirus disease 2019 COVID-19 is an emerging betacoronavirus caused by severe acute respiratory syndrome coronavirus 2 (SARS-CoV-2) \cite{sohrabi_world_2020}. The first infected case was reported in Hubei, a province in the city of Wuhan in China, on December 31, 2019 \cite{guan_clinical_2020}. Infected coronavirus people commonly experience mild to severe respiratory illness, and occasionally death in the most critical cases \cite{sohrabi_world_2020}. The danger of the COVID-19 comparing to the other coronavirus families lies in its asymptomatic and high human-to-human transmission \cite{shereen_covid-19_2020}. Moreover, there are to date no clinical vaccines to prevent the virus and no specific therapeutic protocols to combat this communicable disease \cite{guan_clinical_2020}. The coronavirus epidemic has been confirmed as a global pandemic that has significantly affected the world, not only healthcare systems but also economics, education, transportation, etc. \cite{shereen_covid-19_2020}. As of May 19, 2020, the virus had spread to 215 countries, with more than 4,622,001 confirmed cases and more than 311,916 reported deaths worldwide, including Algeria with 7201 cases and 555 deaths\footnote{www.africacdc.org/download/outbreak-brief-18-covid-19-pandemic-19-may-2020}.
\\\\\indent In the present circumstances, understanding the early transmission dynamics of the novel coronavirus plays important roles in the prevention of its critical impact on human lives and national economies, including in Algeria. Boccaletti et al. \cite{boccaletti_modeling_2020} identified three scientific communities that may cooperate in dealing with the current pandemic: (1) the mathematicians, virologists and epidemiologists who develop sophisticated diffusion models to the specific properties of a given pathogen; (2) the complex systems scientists who study the spread of infections using statistical mechanics and non-linear dynamics methods; and (3) the statisticians and computer scientists who incorporate artificial intelligence and machine learning approaches to develop accurate predictive models. For instance, analyze and develop a time series forecasting model that predicts cases and deaths growth could help governments in decision-making, e.g. take appropriate policy responses, estimate medical equipment (hospital beds, ventilators, protective masks, etc.), and approximate the peak and the end of the epidemic.
\\\\\indent In this perspective, several works have been proposed in recent months in an attempt to understand and forecast the spread of the disease; however, as the COVID-19 is a recent emerging virus, only a few works have been peer-reviewed to date, making it difficult to survey research in analyzing and forecasting coronavirus cases and deaths grow. Nonetheless, significant amounts of prior research have studied and forecasted the spread of similar disease outbreaks such as hepatitis, syphilis, gonorrhoea, seasonal flu, etc. \cite{zhang_applications_2014}. In this context, time series models have long been of interest in the literature to model the spread of epidemic diseases as a series of data points recorded at successive equally spaced points in time \cite{tseng_developing_2019}, as is the case with the coronavirus where the number of cases and deaths are recorded on a daily basis. Time series models typically attempt to predict future outbreak behaviours by modelling and analyzing historical epidemic surveillance data \cite{lu_forecasting_2020}. According to Duarte and Faerman \cite{duarte_comparison_2019}, the most commonly used time series models for forecasting epidemic diseases are AutoRegressive Integrated Moving Average (ARIMA) \cite{box_distribution_1970}, Long Short-Term Memory networks (LSTM) \cite{hochreiter_long_1997}, and \textit{Prophet} \cite{taylor_forecasting_2018}. In this work, we use the \textit{Prophet} model to analyze and forecast the spread of the coronavirus. Note that other time series models are also applicable and that it is out of the scope of this paper to compare them. \textit{Prophet} was proposed by the \textit{Facebook} data science team, who found it to be more effective than classical approaches in the majority of cases \cite{taylor_forecasting_2018}. \textit{Prophet} has been used to forecast a wide range of problems such as cash flow \cite{weytjens_cash_2019}, air pollution \cite{samal_time_2019}, healthcare emergency department indicators \cite{duarte_comparison_2019}, but also in the analysis and forecasting of epidemic diseases, e.g. the influenza epidemic trend \cite{tseng_developing_2019} and seasonal flu activity \cite{lu_forecasting_2020}.
\\\\\indent In this work, we first train a time series \textit{Prophet} model to analyze historical data in Algeria since the beginning of the epidemic and forecast the number of cases and deaths for the coming weeks\footnote{The content of this work is only for research purposes. The results and the predictions must be taken with caution.}. To the best of our knowledge, this is one of the first works that study and forecast the outbreak in Algeria while it is still ongoing. Then, to get a thorough understanding of the properties of the COVID-19, we construct a dataset from different reliable sources and conduct a large-scale data analysis considering 82 countries worldwide. The paper is structured as follows: section \ref{sec3} describes the time series \textit{Prophet} model we use to predict the number of COVID-19 cases and deaths in Algeria, section \ref{sec4} discusses the experiments we performed to evaluate the model and analyze worldwide data, and section \ref{sec5} concludes the paper and proposes future work.

\section{Time series forecasting with \textit{Prophet}}\label{sec3}
\begin{figure}[!thb]
\centerline{\includegraphics[width=0.8\columnwidth]{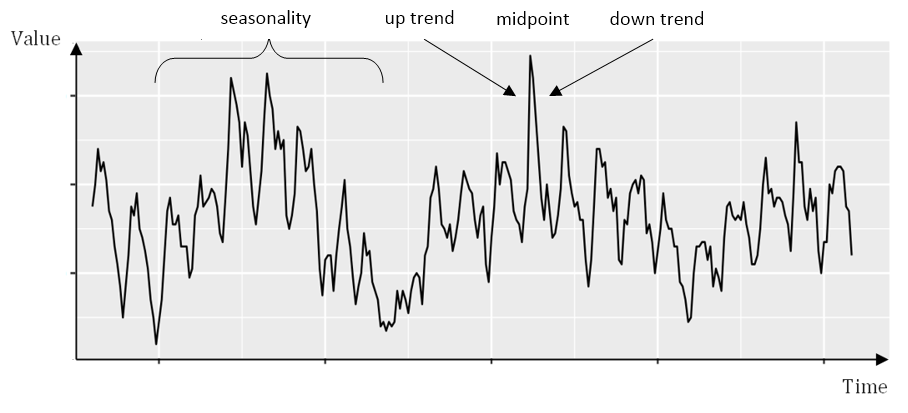}}
\caption{A typical time series curve}
\label{fig0}
\end{figure}

A time series is a series of data points recorded at successive equally spaced points in time, as is the case with the coronavirus where the number of cases and deaths are recorded on a daily basis. Time series are frequently plotted via linear curves (see Fig.~\ref{fig0}), while time series forecasting models aim to predict an increase or decrease in the future values based on previously observed values \cite{chatfield_time-series_2000}.\\

\textit{Prophet} is an open-source time series forecasting software released by \textit{Facebook}, who found it to be more effective than any other approach in the majority of cases \cite{taylor_forecasting_2018}. The \textit{Prophet} model uses a decomposable time series model with three main components combined in Equation \ref{eq1}: trend, seasonality, and holidays. The trend function models non-periodic changes in the time series value, seasonality represents periodic changes (e.g., weekly and yearly seasonality), and the holiday component represents holiday effects which occur on irregular schedules over one or several days. As to the error term, it represents idiosyncratic changes which are not accommodated by the model; the authors make the parametric assumption that the error is normally distributed \cite{taylor_forecasting_2018}. In this paper, we focus on the non-periodic trend function and exclude the seasonality and holiday components. Indeed, the coronavirus is unlikely to be affected by holidays, weekdays, or months. Moreover, we need a minimum of one year to be able to use the yearly seasonality, whereas the virus only emerged six months ago. The use of the \textit{Prophet} model is motivated by its advantages, namely \cite{taylor_forecasting_2018}:

\begin{itemize}
	\item Performance: it gives accurate forecasts on a wide range of problems, including epidemic diseases \cite{tseng_developing_2019}.
	\item Flexibility: it allows to add a multiple periods seasonality according to the trends and the domain knowledge.
	\item Fully automatic: it is not necessary to space measurements, interpolate missing values, or remove outliers.
	\item Speed: it is easy to explore many specifications thanks to the model speed.
	\item Tunable forecasts: it allows changes and assumptions on the forecast thanks to the interpretable parameters.
	\item Implementation: a library that implements the model is available in \textit{Python} and \textit{R} programming languages.
	\end{itemize}

\begin{equation}
y(t) = g(t) + s(t) + h(t) + \epsilon_{t}
\label{eq1}
\end{equation}

Where:
\begin{itemize}
	\item $y(t)$ is the forecast value at time $t$
	\item $g(t)$ is the non-periodic trend at time $t$
	\item $s(t)$ is the periodic (weekly, monthly, yearly) seasonality at time $t$ 
	\item $h(t)$ is the irregular holiday effects at time $t$
	\item $\epsilon_{t}$ is the error term
\end{itemize}

The \textit{Prophet} model has two trend functions $g(t)$ that cover many applications: a saturating growth model which saturates at maximum carrying capacity, and a piecewise linear model that does not exhibit saturating growth \cite{taylor_forecasting_2018}. In this work, we use the saturating growth because, as the number of cases and deaths due to coronavirus, it is characterized by increasing growth in the beginning period and decreasing growth at a later stage, when the forecast values get closer to maximum capacity. For example, to forecast the number of coronavirus cases in a given country, the maximum capacity could be the number of its population where the virus infects all people. The saturating growth is modelled using the logistic growth function given by Equation \ref{eq2}; we discuss its principle in the rest of the section, while further details are provided in \cite{taylor_forecasting_2018}.

\begin{equation}
g(t)=\frac{C(t)}{1+\exp \left(-\left(k+\mathbf{a}(t)^{\top} \boldsymbol{\delta}\right)\left(t-\left(m+\mathbf{a}(t)^{\top} \boldsymbol{\gamma}\right)\right)\right)}
\label{eq2}
\end{equation}

Where:
\begin{itemize}
	\item $g(t)$ is the non-periodic trend at time $t$, and the forecast value in our case
	\item $C(t)$ is the limiting value, the maximum carrying capacity at time $t$
	\item exp is the exponential function, we write it this way for readability reasons
	\item $k+\mathbf{a}(t)^{\top} \boldsymbol{\delta}$ is the growth rate at time $t$
	\item $m+\mathbf{a}(t)^{\top} \gamma$ is the offset parameter at time $t$
\end{itemize}

In Equation \ref{eq2}, if we represent a time series via linear curves (see Fig.~\ref{fig0}), the growth rate $k$ would indicate how steep the curve is, and the offset parameter $m$ would indicate the value of the midpoint where the growth switches from increasing to decreasing \cite{taylor_forecasting_2018}. In the coronavirus case, the time $t$ indicates the number of days since the first recorded case, $k$ corresponds to the virus spread rate, and $m$ indicates the value of the midpoint where the number of daily new cases/deaths starts to decrease. To estimate the parameters $k$ and $m$ at time $t$, the \textit{Prophet} model learns the values that best fit the data during the training process \cite{taylor_forecasting_2018}. The aim is to fit the model on historical training data since the beginning of the epidemic and use it to predict the number of cases and deaths in the coming weeks. As to the maximum carrying capacity $C$ at time $t$, we calculate its value using Equation \ref{eq3}. According to experts, one COVID-19 patient infects between two and three other people\footnote{www.ecdc.europa.eu/en/covid-19/questions-answers}; therefore, we assume that the maximum carrying capacity corresponds to the case where each patient infects three persons. In the next section, we discuss the experiments we performed to train and evaluate the time series model on coronavirus data in Algeria and analyze worldwide COVID-19 data.

\begin{equation}
C(t) = t^{3}
\label{eq3}
\end{equation}

Where: 
\begin{itemize}
	\item $t$ is the number of days since the first recorded coronavirus case
	\item 3 is the average number of people infected by a coronavirus patient
\end{itemize}

\section{Experimentation and analysis}\label{sec4}
All the experiments and data analysis we performed are based on the reliable \textit{Our World in Data} website \cite{owidcoronavirus}, which is updated daily. It provides official worldwide COVID-19 statistics according to several criteria: by country, cases, deaths, tests performed, mortality risk, and government policy responses. The results, figures, and statistics presented in this section have been reported at the time of writing (19/05/2020) while the epidemic is still ongoing. The rest of this section discusses the experimentation we conducted after several data preprocessing steps using \textit{Python} programming language and \textit{Jupyter Notebook}\footnote{https://jupyter.org}. The experiments are carried out in two stages: evaluate the time series model in predicting the spread of the virus in Algeria, and data analysis considering data from 82 countries worldwide for a better understanding of the pandemic.

\subsection{Time series forecasting: Algeria}
In a series of experiments, we analyzed historical data, trained, and evaluated the \textit{Prophet} model in forecasting the spread of the virus in Algeria. Note that Algeria is a case study and that the model can be used for any other country.

\begin{figure}[!thb]
\centerline{\includegraphics[width=\columnwidth]{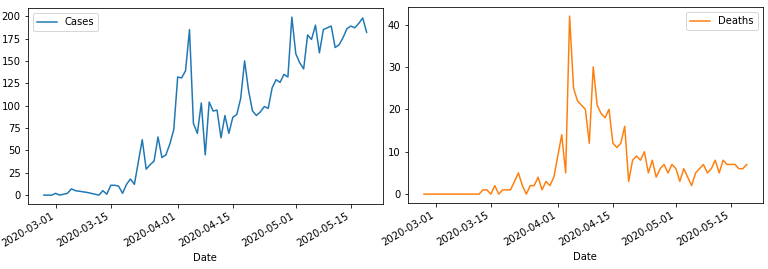}}
\caption{Daily new coronavirus cases and deaths in Algeria}
\label{fig1}
\end{figure}

First, Fig.~\ref{fig1} gives the number of daily new cases and deaths in Algeria, from the first case to date. The daily death curve suggests that the epidemic reached its peak in early April when Algeria reported an unfortunate record of 42 deaths in a single day. The death curve has been steadily decreasing since then and then stabilizing during the first weeks of May with less than ten deaths a day. As to the daily case curve, the first case was recorded on February 26, 2020. The curve has been steadily increasing since then until it reached its first peak in early April at the same time as the death peak when Algeria reported 185 cases in a single day. Later, unlike the death curve, the daily case curve first decreased in April and then peaked again in early May with a new record of 199 cases in a single day. The rise in the number of cases is certainly related to the increase in the number of COVID-19 tests performed. Indeed, with more labs across the country, the head of the Algerian Pasteur institute stated that the number of tests increased to an average of 400 tests per day in early May compared to 200 tests at the beginning of the epidemic\footnote{www.aps.dz/sante-science-technologie/104593-covid-19-hausse-sensible-du-nombre-de-tests-quotidiens-dans-les-laboratoires-de-l-ipa\label{fn1}}. Later, the case curve has been stabilizing in the first weeks of May with less than 200 cases a day.

\begin{figure}[!thb]
\centerline{\includegraphics[width=\columnwidth]{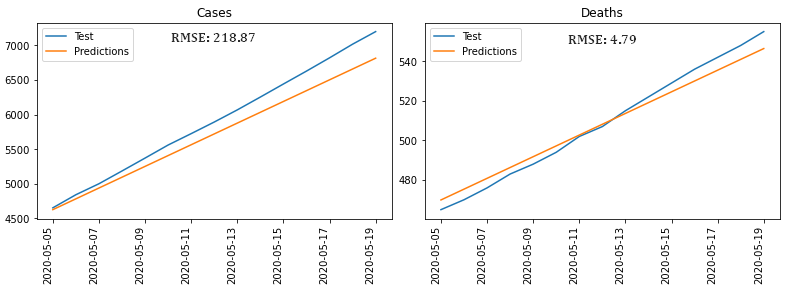}}
\caption{Evaluation results over the past 15 days}
\label{fig2}
\end{figure}

Second, we used the total number of coronavirus cases and deaths per day in Algeria from 26 February to 04 May 2020 to train the time series forecasting model described in section \ref{sec3} and evaluate it on the remaining past 15 days, from 05 to 19 May 2020. The case model and the death model are trained separately; the aim is to fit the model on historical data since the beginning of the epidemic and use it to predict the number of cases and deaths in the coming weeks. Fig.~\ref{fig2} shows the evaluation results corresponding to the predicted and actual test values, and Equation \ref{eq4} gives the Root Mean Squared Error (RMSE) measure, which is commonly used to evaluate the performance of time series models \cite{tseng_developing_2019}. Fig.~\ref{fig2} shows that the model accurately predicts the number of cases and deaths with low RMSE scores of 218.87 and 4.79, respectively. Note that the error score for the number of predicted cases is higher than the score for the number of deaths only because the number of cases is highest. Moreover, the figure indicates that the case and death functions the model has learned are quasi-linear from 05 to 19 May 2020, as the actual case and death curves. We calculated a linear approximation function from the two latter curves which are as follows: Y1 = 157 X1 + 4465 for cases, and Y2 = 5 X2 + 464 for deaths; where 4465 and 464 are the number of cases and deaths respectively at the beginning of the forecast. The functions suggest that the country recorded approximately the same daily number of cases and deaths during this period, i.e. 157 cases and 5 deaths. These results are consistent with our observations in Fig.~\ref{fig1}, where the case and death curves have stabilized in early May.

\begin{equation}
RMSE = \sqrt{(\frac{1}{n})\sum_{i=1}^{n}(y_{i} - x_{i})^{2}}
\label{eq4}
\end{equation}

Where:
\begin{itemize}
	\item $n$ is the number of days
	\item \textit{{y}$_{i}$} is the predicted value for the day $i$
	\item \textit{{x}$_{i}$} is the actual test value for the day $i$
\end{itemize}

\begin{figure}[!thb]
\centerline{\includegraphics[width=\columnwidth]{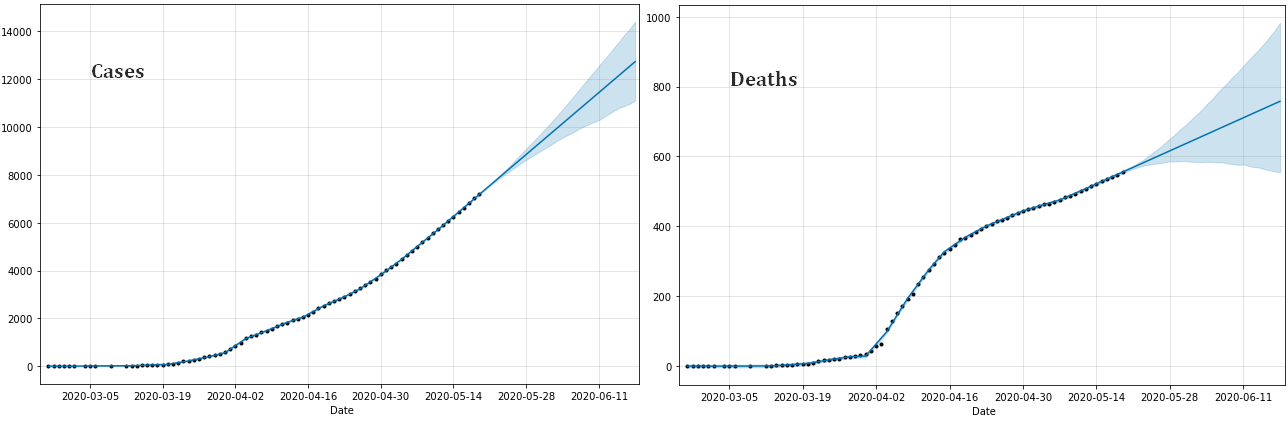}}
\caption{Forecast for next month starting on 20/05/2020}
\label{fig3}
\end{figure}

Finally, we retrain the time series model on all data available to date to forecast the total number of coronavirus cases and deaths for next month, starting on 20/05/2020 (see Fig.~\ref{fig3}). Note that the case model and the death model are trained separately. The case curve indicates that the number of cases, currently 7201, is expected to increase by 4800 to reach 12000 cases by mid-June, which corresponds to an average of 185 new cases a day. On the other hand, the death curve suggests that the number of deaths, currently 555, is expected to increase by 195 to reach 750 deaths by mid-June, which corresponds to an average of 8 new deaths a day. Therefore, we note that both case and death curves do not show a downward trend for the coming month. Indeed, as shown in Fig.~\ref{fig1}, the number of daily new recorded cases/deaths has been continuously decreasing and increasing again in early May, making it difficult for the time series model to detect a downward trend and predict the end of the epidemic. Furthermore, time series forecasting models only take the previously reported cases and deaths as input to predict the future numbers, without any knowledge of the properties of the COVID-19. In the next section, we include and analyze other factors that may contribute to accelerate/slow the spread of the virus.

\subsection{Data analysis: Countries of the world}
To get a better understanding of the properties of the COVID-19, we conducted a large-scale data analysis considering 82 countries worldwide. In this section, we describe the dataset we constructed as well as the large-scale data analysis.

\subsubsection{Dataset}
First, Table~\ref{tab1} describes the dataset we constructed from different reliable sources, where each row represents a country, and the columns represent geographic, climate, healthcare, economic, and demographic factors that may contribute to accelerate/slow the spread of the coronavirus \cite{owidcoronavirus}. The dataset is publicly available and updated monthly with the latest numbers of COVID-19 cases, deaths, tests.\footnote{https://github.com/SamBelkacem/COVID19-Algeria-and-World-Dataset}. Note that we selected only the main factors for which we found data and that other factors can be used. For the rest of the paper, we will use the term "features" instead of "factors". Indeed, we believe that leveraging the features in supervised learning algorithms is one of the future steps in understanding and predicting the number of COVID-19 cases and deaths. As indicated in section \ref{sec4}, we retrieved all data from the reliable \textit{Our World in Data} website \cite{owidcoronavirus}, except for data on continents\footnote{www.kaggle.com/statchaitya/country-to-continent}, geographic-coordinates\footnote{www.kaggle.com/eidanch/counties-geographic-coordinates}, temperatures\footnote{www.kaggle.com/berkeleyearth/climate-change-earth-surface-temperature-data}, and share of the population over 65 years old\footnote{https://data.worldbank.org/indicator/SP.POP.65UP.TO.ZS}. The assumptions for the different feature categories are as follows:

\begin{table}[!thb]
\caption{Data description}
\label{tab1}
\includegraphics[width=\columnwidth]{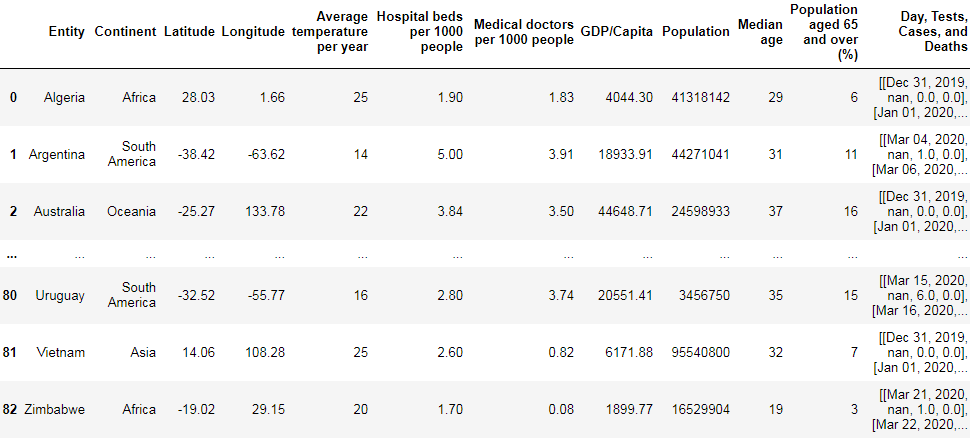}
\end{table}

\begin{itemize}
	\item Geography: some continents/areas may be more affected by the disease
	\item Climate: cold temperatures may promote the spread of the virus
	\item Healthcare: lack of hospital beds/doctors may lead to more human losses
	\item Economy: weak economies (GDP) have fewer means to fight the disease
	\item Demography: older populations may be at higher risk of the disease 
\end{itemize}

Second, the last column of Table~\ref{tab1} gives, in list forms, the total number of coronavirus cases, deaths, and tests performed per day for each of the 83 countries in the dataset, from the first day of the epidemic to date. Countries that do not provide any information on the number of tests performed are not included in the dataset. As to Algeria, we added the daily number of tests performed according to the numbers provided by the Algerian Pasteur institute: 200 tests per day since the beginning of the epidemic and 400 tests per day in May\textsuperscript{\ref{fn1}}. The last column is then split into several columns and rows, resulting in 15 columns and 7050 rows in the dataset. Moreover, to fill missing values in some dates (112 missing values in cases, 112 in deaths, and 4102 in tests), we propagated values from the previous days to the following days. For example, if a given country performed 500 tests on a given day, and a value is missing for the next day, we assume the country performed as many tests as the previous day.\\

\begin{figure}[!thb]
\centerline{\includegraphics[width=0.93\columnwidth]{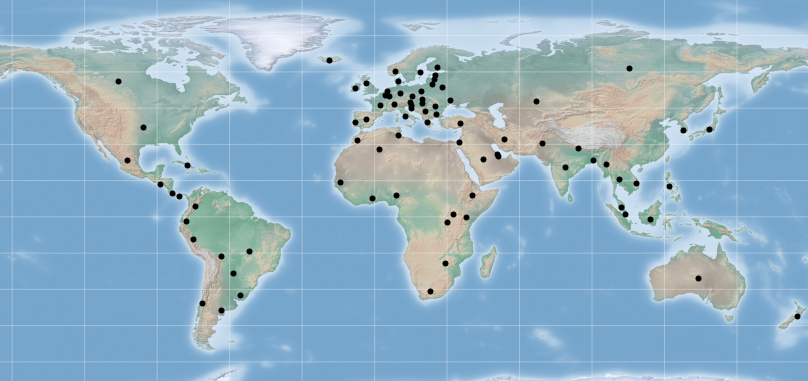}}
\caption{Countries in the dataset by geographic coordinates}
\label{fig5}
\end{figure}

Finally, Fig.~\ref{fig5} shows the geographical coordinates of all countries in the dataset where each data point corresponds to the latitude and longitude of a given country. The 82 countries in the dataset are distributed over the six continents of the world as follows: 32 countries in Europe, 21 in Asia, 12 in Africa, 9 in South American, 7 in North America, and 2 in Oceania. The imbalanced distribution of countries by continent is due to several factors\footnote{https://worldpopulationreview.com/countries/continent-with-the-most-countries}: (1) North America is represented with only seven countries because, except a few large countries, e.g. United States, Canada, and Mexico, most other countries are small islands where the virus spread moderately, e.g. Bahamas, Haiti, Saint Lucia, etc. \cite{shereen_covid-19_2020}; (2) South America and Oceania are represented with only 9 and 2 countries respectively because these continents have fewer countries with 12 and 14 countries respectively; (3) Africa is represented with only 12 countries out of 48 because most countries do not provide any information on the number of coronavirus tests performed; such countries were not included in the dataset; and (4) Europe and Asia are highly represented with 32 countries and 21 countries out of 44 and 32 respectively because, at the time of writing (19/05/2020), the virus first appeared in China, then spread widely in Asia and Europe \cite{guan_clinical_2020}.

\subsubsection{Data analysis}

\begin{figure}[!thb]
\centerline{\includegraphics[width=0.95\columnwidth]{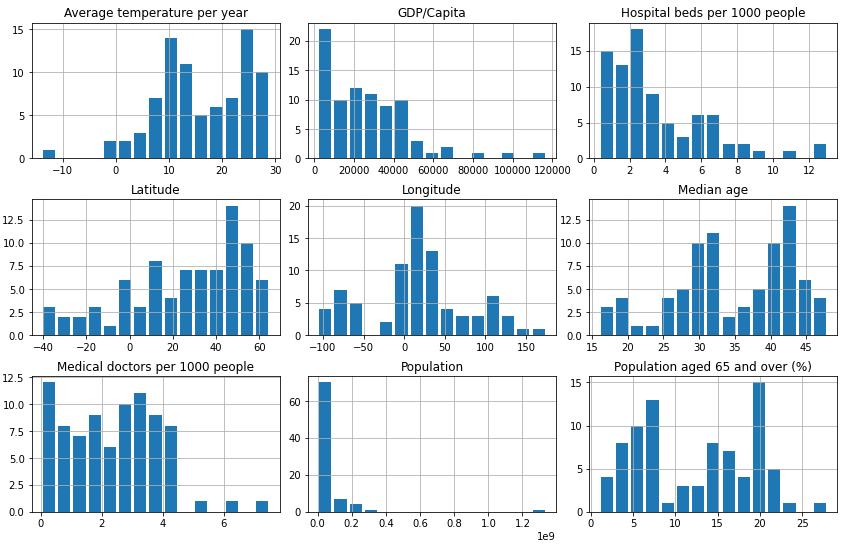}}
\caption{Feature distribution in all countries}
\label{fig6}
\end{figure}

First, to analyze data of Table~\ref{tab1} and visualize the features that may contribute to accelerate/slow the spread of coronavirus, we show in Fig.~\ref{fig6} the distribution of feature values across all countries in the dataset. The x-axis corresponds to the feature value, and the y-axis corresponds to the number of countries with that value. Due to lack of space, we discuss only the most important findings. The figure shows that the geographic coordinates range from the extreme north in Iceland (latitude of 64.96) to the extreme south in New Zealand (latitude of -40.9), and from the extreme east in New Zealand (longitude of 174.89) to the extreme west in Canada (longitude of -106.35). The assumption for the geographical features is that some continents or geographic areas may be more affected by the disease. The average temperature a year ranges from -14\textcelsius{} in Denmark to 29\textcelsius{} in Senegal, Nigeria, and Qatar. Note 23 cold countries that have an average temperature below 10\textcelsius{} and 22 warm countries that have an average temperature above 24\textcelsius{}. The assumption for this feature is that warm temperatures may slow the virus. The number of hospital beds per 1000 people ranges from 0.3 in Ethiopia, Senegal, and Nepal to 13.05 in Japan, where 28 countries have less than two beds per 1000 people. The assumption for this feature is that the fewer hospital beds available, the more overwhelmed the hospitals, and the more human losses. The number of medical doctors per 1000 people ranges from 0.02 in Ethiopia to 7.52 in Cuba, where 20 countries have less than one doctor per 1000 people. The assumption for this feature is that the fewer doctors, the fewer medical staff to manage infected patients, and the more human losses. The GDP/Capita ranges from 1697.71 USD in Uganda to 116936 USD in Qatar, where 22 countries have a GDP/Capita of less than 10000 USD. The assumption for this feature is that low-income countries may have fewer means to fight the disease. The population size ranges from 341284 inhabitants in Iceland to 1,386,395,000 inhabitants in China, where 12 countries have more than 100 million inhabitants. The assumption for this feature is that the number of cases and deaths may be higher in populous countries. Finally, the median age ranges from 16 years old in Uganda to 48 years old in Japan, where the median age is over 40 years old in 31 countries. As to the share of the population over 65, it ranges from 1\% in Qatar to 28\% in Japan, where the share is over 15\% in 33 countries. The assumption for these features is that older populations are at high risk of contracting and dying from the disease.

\begin{figure}[!thb]
\centerline{\includegraphics[width=\columnwidth]{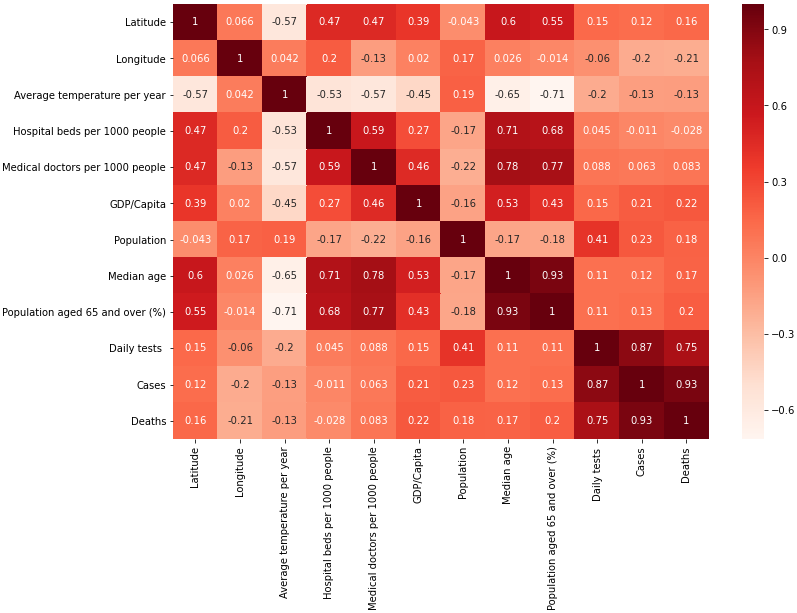}}
\caption{Feature correlation in all countries as of 19/05/2020}
\label{fig7}
\end{figure}

Second, to find relationships in the dataset between the features and the number of cases and deaths in the 83 countries to date, we computed in the heatmap of Fig.~\ref{fig7} the Pearson correlation coefficient given by Equation \ref{eq5} \cite{lawrence_concordance_1989}. The latter measures the statistical relationship between two continuous variables and has a value between +1 and -1, where 1 is a total positive linear correlation, 0 is no linear correlation, and -1 is total negative linear correlation. Due to lack of space, we discuss only the most important correlations. The heatmap indicates a strong positive correlation (0.87) between the number of tests performed and the number of cases, emphasizing the importance of testing to detect infected patients, isolate and promptly treat them, and limit the spread of the virus. The median age and the share of the population over 65 are positively correlated with the number of deaths (0.17 and 0.2 respectively), which confirms that older populations are at high risk of death from the disease. The population size is negatively correlated with the number of medical doctors and beds per 1000 people (-0.22 and -0.17 respectively), which suggests that populous countries may have less healthcare means to manage infected patients. The heatmap also indicates a negative correlation between the average temperature a year and the number of cases (-0.13), which suggests that warm temperatures may slow the virus. The number of hospital beds per 1000 people is negatively correlated with the number of deaths (-0.024), which confirms that the more hospital beds available, the less overwhelmed the hospitals, and the less human losses.

\begin{equation}
  r = \frac{\sum_{i=1}^{n}(x_i-\bar{x})(y_i-\bar{y})}{%
	    \sqrt{\sum_{i=1}^{n}(x_i-\bar{x})^2}\sqrt{\sum_{i=1}^{n}(y_i-\bar{y})^2}}
\label{eq5}
\end{equation}

Where:
\begin{itemize}
	\item $\bar{x}$ is the mean of the variable $x$
	\item $\bar{y}$ is the mean of the variable $y$
	\item $n$ is the number of data pairs between $x$ and $y$
\end{itemize}

Finally, the heatmap of Fig.~\ref{fig7} shows a surprising but expected positive correlation between the GDP/Capita and the number of cases and deaths (0.21 and 0.22 respectively). Indeed, the highest case and death tolls have been reported in wealthier countries in Europe and the United States. Experts have suggested many reasons\footnote{www.voanews.com/covid-19-pandemic/why-covid-19-hit-high-income-countries-harder}: (1) the climate which is colder in the northern countries of the world (-0.57 correlation between temperature and latitude); (2) the testing gap between high and low-income countries (0.15 correlation between GDP/Capita and the number of tests performed); (3) the high connectivity to China which is the country of origin of the coronavirus; and (4) the population density which is higher in developed countries. Nonetheless, the correlation heatmap indicates that the GDP/Capita is positively correlated with the number of tests performed, hospital beds, and medical doctors (0.15, 0.27, and 0.46 respectively). High-income countries undoubtedly have more human and material means to fight the disease despite the extent of the epidemic on their soil. It is lastly important to note that the correlations given in Fig.~\ref{fig7} only quantify the strength of the relationship between features, which does not always indicate causation. Investigate feature correlation is a first step to explore relationships between COVID-19 and external factors that contribute to accelerate/slow the spread of the virus. Further analysis is required to make a proper causal inference.

\section{Conclusion}\label{sec5}
In this work, we first trained a time series \textit{Prophet} model to analyze and forecast the number of COVID-19 cases and deaths in Algeria. Then, to better understand the spread of the virus, we constructed a dataset from different reliable sources and conducted a large-scale data analysis considering 82 countries worldwide. The evaluation results show that the time series model accurately predicts the number of cases and deaths in Algeria with low RMSE scores of 218.87 and 4.79 respectively, while the forecast suggests that the total number of cases and deaths are expected to increase in the coming weeks. Moreover, the worldwide data-driven analysis reveals several correlations between the increase/decrease in the number of cases and deaths and other factors that may contribute to accelerate/slow the spread of the virus, such as geographic, climatic, health, economic, and demographic factors. However, this is a first attempt to analyze and forecast the epidemic while it is still ongoing. Further analysis is required to get a thorough understanding of the properties of the coronavirus.
\\\\\indent For future work, we intend to include other features that contribute to accelerate/slow the spread of the virus, especially regarding policy responses to the COVID-19, e.g. restrictions on public gatherings, international and domestic travel, etc. Furthermore, to better understand and forecast the spread of the virus, we plan to extend the data with the latest numbers and train a supervised model that predicts the number of cases and deaths for any country.

\bibliographystyle{unsrt} 
\bibliography{Mabib}
\end{document}